\documentclass{article}

\usepackage[english]{babel}
\usepackage{natbib}

\usepackage[dvips]{graphicx}
\usepackage{amsfonts}
\usepackage[mathscr]{eucal}
\usepackage{epsfig}

\usepackage{amssymb,amsmath}

\begin{document}

\large

\title{Role of network connectivity in
intercellular calcium signaling}

\author{I.V. Dokukina$^{a,c}$, M.E.Gracheva$^{b}$, E.A.Grachev$^{c}$, and
J.D.Gunton$^{a}$ \vspace{12pt} \\ $^{a}$Department of Physics, Lehigh
University, PA  18015, USA \\ $^{b}$University of Illinois at Urbana-Champaign,
\\ Beckman Institute for Advanced Science and Technology, Urbana, \\IL 61801,
USA \\ $^{c}$Faculty of Physics, Moscow State University, Moscow, Russia}

\maketitle

\begin{abstract}
It is important to understand the coordinated performance of cells in tissue.
One  possible mechanism in this coordination involves intracellular $\rm
Ca^{2+}$ signaling. The topology of intercellular connections in tissue should
also play an important role in this process. It is most relevant for plane
tissues, in which the interaction between cells is due to gap junctions
(epithelium, blood vessels). We  demonstrate the importance of the topology of
intercellular connectivity by investigating the properties of a model of $\rm
Ca^{2+}$ signaling for a small number of connected cells.
\end{abstract}


\vspace{12pt} \textbf{E-mail addresses of all authors:}\\ Irina V. Dokukina:
dokukina@gmail.com
\\Maria E. Gracheva: gracheva@uiuc.edu\\ Eugene A. Grachev:
grachevea@gmail.com; grace@cmp.phys.msu.ru\\ James D. Gunton: jdg4@lehigh.edu

\vspace{12pt} \textbf{Corresponding author:} Irina V. Dokukina. Permanent
address: Moscow State University, Faculty of Physics, building 2, room 2-40a,
Leninskie Gory, Moscow, 119992, Russian Federation. Phone: 7-(495)-939-4178.
E-mail address: dokukina@gmail.com.

\section{Introduction}

Calcium ions ($\rm Ca^{2+}$) are one of the universal regulators of
various intra- and intercellular processes \cite{berridge-94}. In
the presence of connections between cells, when each cell is a piece
of tissue, $\rm Ca^{2+}$ signal can be passed from cell to cell as
intercellular $\rm Ca^{2+}$ wave \cite{berridge-98}. The importance
of intercellular $\rm Ca^{2+}$ signaling (among other secondary
messengers) in the coordinated performance of cells in tissues is
currently under investigation.  These waves have been seen in a
variety of cell types \cite{boitano-92}, \cite{loessburg-93},
\cite{cornell-90}, \cite{dani-92}. Intercellular $\rm Ca^{2+}$ waves
can be initiated by a focal mechanical, electrical, or hormonal
stimulus. Often such waves travel from cell to cell in an
oscillatory manner, thus forming periodic intercellular calcium
waves \cite{clair-01}, \cite{evans-99}.

Intercellular $\rm Ca^{2+}$ signaling may occur through different pathways, for
example through paracrine or junctional routes. In recent study
\cite{iacobas-06} it was shown for astrocytes that propagation of intercellular
$\rm Ca^{2+}$ waves in glia is essentially determined by both gap junctions
between cells and paracrine signaling by ATP, other nucleotides and/or their
metabolites. A detailed theoretical and experimental study of calcium
oscillations in astrocytes doublets and linearly coupled triplets is given by
Ullah et al. \cite{ullah-06}. A model of intercellular calcium waves based on
the action of calcium as a second messenger in paracrine signal transduction
via calcium sensing receptor was proposed by Gracheva and Gunton
\cite{gracheva-03} and extended for a linear chain of cells by Kepseu and Woafo
\cite{kepseu-06}.

Liver, blood vessels, brain and epithelium are very good examples of tissues in
which cells are connected with each other by gap junctions. It was shown by
experiments and modeling that mechanically stimulated intercellular $\rm
Ca^{2+}$ waves in epithelial, glial and endothelial cells can result from the
diffusion of inositol 1,4,5-trisphosphate ($\rm IP_{3}$) through gap junctions
\cite{leybaert-98}, \cite{sanderson-94}, \cite{sneyd-98}. Although $\rm
Ca^{2+}$ can move through gap junctions, a cellular $\rm Ca^{2+}$ wave is the
cellular response to a diffusing of $\rm IP_{3}$, passing between cells via gap
junctions and consequently participating in the release of $\rm Ca^{2+}$ ions
from the intracellular stores. However, in contrast with these cell types, the
amount of $\rm IP_{3}$ diffusing through the gap junctions in connected
hepatocytes is insufficient to induce a $\rm Ca^{2+}$ response \cite{clair-01},
\cite{clair-03}, \cite{dupont-00b}, \cite{tordjmann-97}. Thus, models based on
the passive diffusion of $\rm IP_{3}$ between adjacent cells through gap
junctions \cite{sneyd-98} cannot account for apparent intercellular $\rm
Ca^{2+}$ waves in hepatocytes. Hoefer \cite{hofer-99} has proposed an
alternative mechanism of intercellular calcium waves in hepatocytes, based on
the diffusion of cytosolic calcium through gap junctions. Stochastic versions
of both the \cite{dupont-00b} and \cite{hofer-99} models are given by Gracheva
et al. \cite{gracheva-01}. Clair et al. \cite{clair-03} have also shown that an
agonist receptor gradient could be responsible for the apparent undirectional
$\rm Ca^{2+}$ waves observed in multiplets or in perfused intact livers. In
this case, the main factor influencing $\rm Ca^{2+}$ wave propagation in
hepatocytes is the cellular distribution of agonist receptors.

In addition, Tsaneva-Atanasova et al. \cite{tsaneva-05} have shown for
pancreatic acinar cells that intercellular calcium diffusion is necessary and
sufficient to synchronize $\rm Ca^{2+}$ oscillations in neighboring cells,
whereas the function of intercellular $\rm IP_{3}$ diffusion is to provide the
propagation of oscillating $\rm Ca^{2+}$ waves in tissue.

In all the above mentioned models the junctional permeability was chosen to be
constant. However, since calcium ions can close gap junctions
\cite{bennett-92}, \cite{saez-93}, this would appear to pose an obstacle to the
propagation of intercellular $\rm Ca^{2+}$ waves via gap junctions.
Fortunately, this apparent paradox can be resolved by considering the temporal
aspects of these responses. It is known that extended exposure to high
concentrations of $\rm Ca^{2+}$ ($> 10~\rm \mu M$) results in the closure of
gap junctions \cite{bennett-92}, \cite{saez-93}. It also appears that
physiological concentrations of $\rm Ca^{2+}$ can reversibly close gap
junctions but this response takes about $30~\rm s$ \cite{lazrak-94}. Thus,
because $\rm Ca^{2+}$ waves propagate faster than the proposed closure times
\cite{sanderson-90}, the two possible intercellular messengers ($\rm Ca^{2+}$
and $\rm IP_{3}$) \cite{saez-89}, \cite{niessen-00} can diffuse to adjacent
cells to propagate the wave before $\rm Ca^{2+}$ closes the gap junctions.
Therefore, it would seem that a more accurate description of the gap junction
permeability requires taking into account its dependence on the cytosolic
calcium concentration, rather than treating it as a constant as in the earlier
models \cite{sneyd-98}, \cite{dupont-00b}, \cite{hofer-99}, \cite{gracheva-01}.

We believe that an important issue in modeling intercellular
calcium signaling is to account for the cellular topology of the
tissue. As a rule, flat tissues are modeled in the form of a
two-dimensional square grid \cite{sneyd-94}, \cite{hofer-02}.
However, in reality, the structure of even flat tissues is
significantly more complicated. Even small areas of tissue can
exhibit different topological types of cell connections between
each other \cite{evans-99}, \cite{niessen-00}. In addition, in the
case of local stimulation in which only one cell (not the entire
tissue) is stimulated, it is important to know which cell is
stimulated (for the same structure of cell connections). Thus, the
aim of our study is to investigate the relation between the
topology of cellular connectivity in flat tissues and the
corresponding intercellular calcium signaling.

Since gap junctions ensure a strict topology of intercellular connections, we
take into account in our study only tissues in which cells are connected with
each other by gap junctions and the main mechanism of intercellular $\rm
Ca^{2+}$ wave propagation is a diffusion of a second messenger through these
gap junctions. We do not exclude the possible influence of extracellular
pathways on the intercellular calcium signaling. However, for the sake of
clarity and simplicity, in this work we omit taking into account any second
messenger diffusion through the extracellular matrix, which could be relevant,
for example, in astrocytic networks \cite{iacobas-06}. In our model, we take
into account only the diffusion of $\rm IP_{3}$ through gap junctions as the
main mechanism of $\rm Ca^{2+}$  wave propagation in tissue \cite{sneyd-98},
\cite{tsaneva-05}, \cite{sanderson-90}. Also we include in our model the
dependence of gap junctional conductance on concentration of cytosolic $\rm
Ca^{2+}$ \cite{lazrak-94}.

\section{The model description}

We use the minimal model \cite{goldbeter-90} as the basis of our model of
intercellular $\rm Ca^{2+}$ signaling. In this model \cite{goldbeter-90} the
total constant entry of $\rm Ca^{2+}$ into the cytosol is
$\nu_{0}+\nu_{1}\beta$. This includes the influx $\nu_{0}$ from the
extracellular medium and the $\rm IP_{3}$-stimulated $\rm Ca^{2+}$ release
$\nu_{1}\beta$ ($\beta$ is the degree of saturation of the $\rm IP_{3}$
receptor). It is more convenient for us to include in the model an explicit
dependence on the $\rm IP_{3}$ concentration in cell. Therefore we replace the
$\rm IP_{3}$-stimulated $\rm Ca^{2+}$ release $\nu_{1}\beta$ in
\cite{goldbeter-90} by its linear dependence on $\rm IP_{3}$:
$k_{inIP_{3}}IP_{3}$.

We also include gap junction tunneling between nearby cells. In order to model
$\rm Ca^{2+}$ wave propagation the $\rm IP_{3}$ stimulus (which mimics the
local application of a stimulus) is applied only to the first cell. We model
this stimulus induced $\rm IP_{3}$ as following function of time for the first
cell:
\begin{equation}
IP_3^{(stimulus)}=\left\{ \begin{array}{l} IP_3^{(max)}\cdot \frac{t}{t_1}, 0<t<t_1,\\
IP_3^{(max)}, t_1\leq t\leq t_2,
\\ 0, t>t_2,
 \end{array}\right.
\end{equation}
where $t$ -- time, $t_1$ -- time moment by which $IP_3$  reaches its maximal
value $IP_3^{(max)}$; $t_2$ - time moment at which $\rm IP_{3}$ stimulation is
interrupted. Thus, we formulate the rate of $\rm IP_{3}$ synthesis $J_{synt}$
as follows:
\begin{equation}
J_{synt}=k_{rate} IP_3^{(stimulus)}.
\end{equation}
Modeling the agonist generated $\rm IP_{3}$ concentration by such function is
more convenient numerically than, for example, using a square impulse. In
addition, we add to the model a term describing the degradation of $\rm IP_{3}$
\cite{sneyd-94} in each of the cells:
\begin{equation}
J_{deg}=\frac{V_{p}IP_{3}k_{p}}{k_{p}+IP_{3}}, \label{eq3}
\end{equation}
where $k_p$ is the half maximal rate of $IP_3$ degradation \cite{sneyd-94}.
Without this term, the calcium levels reach unphysiological values in the
asymptotic time limit \cite{sanderson-90}. The values of all the parameters are
given in Table \ref{tabl1article}.

Lazrak et al. \cite{lazrak-94} have shown that during the fast increase of
cytosolic $\rm Ca^{2+}$  concentration in each of the coupled adjacent cells
(from $0.2~\rm \mu M$ to $1~\rm \mu M$), the gap junctional conductance
decreases from $100~\%$ to $0~\%$. During the subsequent return of the level of
cytosolic $\rm Ca^{2+}$ to its normal value (about $0.2~\rm \mu M$), the gap
junctional conductance recovers with practically the same rate as its previous
decrease (see Fig.~3 in \cite{lazrak-94}). This means that during each
oscillation of cytosolic $\rm Ca^{2+}$ in one of two adjacent cells, the gap
junctions between these cells are transiently closed. However, this closure
does not occur instantaneously, but during an interval of about $30~\rm s$.
Relying on these data (Fig.~3 in \cite{lazrak-94}), we model the fast changes
in the gap junctional conductance $\gamma_{ij}$ between adjacent cells $i$ and
$j$ by the following function:
\begin{equation}
\gamma_{ij}=\frac{k_{ij}^n}{Ca_{cytMax(ij)}^n+k_{G}^n},
\label{eq4}
\end{equation}
where $Ca_{cytMax(ij)}$ is the larger of the two cytosolic calcium
concentrations in cells $i$ and $j$. Thus, $\gamma_{ij}=1$ corresponds to a
$100~\%$ conductive gap junction between cell $i$ and $j$. For other parameters
see Table \ref{tabl1article}. The gap junctional conductance in Fig.~3 in
\cite{lazrak-94} decreases very rapidly, even with a small increase of
cytosolic $\rm Ca^{2+}$. Therefore we choose $n=5$ in the function (\ref{eq4})
(see Fig.~1). The diffusion of $\rm IP_{3}$ provides the mechanism of
intercellular $\rm Ca^{2+}$ wave propagation in our model.

Thus, the evolution of our system is described by the following
differential equations (cf. \cite{goldbeter-90}, for parameter
values see Table \ref{tabl1article}):
\begin{equation}
\frac{{dCa_{cyt(i)} }} {{dt}} = k_{in}+k_{inIP_{3}}IP_{3(i)} -
k_{out}Ca_{cyt(i)} + \nonumber
\end{equation}
\begin{equation}
+(k_{pool}Ca_{pool(i)}+k_{rel}\frac{Ca_{pool(i)}^2}{K_{1}^{2}+Ca_{pool(i)}^{2}}\frac{Ca_{cyt(i)}^{4}}{K_{2}^{2}+Ca_{cyt(i)}^{4}})
-  k_{serca}\frac{Ca_{cyt(i)}^2}{K_{3}^2+Ca_{cyt(i)}^2},
\label{eq1}
\end{equation}
\begin{equation}
\frac{{dCa_{pool(i)} }} {{dt}} =
k_{serca}\frac{Ca_{cyt(i)}^2}{K_{3}^2+Ca_{cyt(i)}^2} -  \nonumber
\end{equation}
\begin{equation}
-k_{rel}\frac{Ca_{pool(i)}^2}{K_{1}^{2}+Ca_{pool(i)}^{2}}\frac{Ca_{cyt(i)}^{4}}{K_{2}^{2}+Ca_{cyt(i)}^{4}}-k_{pool}Ca_{pool(i)},
\label{eq2}
\end{equation}
\begin{equation}
\frac{dIP_{3(i)}}{dt}=J_{synt}-J_{deg}+k_{gap}\gamma_{ij}(IP_{3(j)}-IP_{3(i)})+k_{gap}\gamma_{ik}(IP_{3(k)}-IP_{3(i)}),
\label{eq5}
\end{equation}
where the indices $i$, $j$ and $k$ denote the cell number, and
$k_{gap}$ is the coefficient of gap junctional communication
 for $\rm IP_{3}$ (see Table \ref{tabl1article}).

Thus, the index triplets $(ijk)=(120)=(210)$ correspond to just two coupled
cells with only $k_{12}=k_{21} \neq 0$. The case in which $(ijk)=(100)$,
$Ca_{cytMax(ij)}=Ca_{cyt(1)}$ and all junctional coefficients are equal to zero
corresponds to just one cell. This is equivalent to equations for intracellular
$\rm Ca^{2+}$ signaling. It is also worth noting that in the case of index
pairs $(ijk)=(123)=(213)=(312)$ there are three possible different
configurations of three coupled cells.  If $k_{13}=0$, we have a
one-dimensional chain of cells, with the first cell interacting only with the
second cell, the second cell interacting with both the first and third cells
and the third cell interacting only with the second cell (see Fig.~2~A). The
same linear chain with other orderings of the cells is shown in Fig.~2~B. If
none of the junctional coefficients vanish, each cell interacts with two other
cells (see Fig.~2~C). The right side of Fig.~2 shows graphs that correspond to
these three configurations.

The system can be easily generalized to the case of more than three cells. For
example, all the configurations in two dimensions of four coupled cells are
shown in Fig.~3. All these are biologically realistic in two dimensions.
Examples of real cellular configurations in vitro are shown in Fig.~1 and 2 in
\cite{niessen-00} and Fig.~9 and 11 in \cite{evans-99}. The system of cells 1,
2, 3, 7 in Fig.~1a in \cite{niessen-00} (if the rest of the cells are removed)
corresponds to the topological configuration in Fig.~3~A (cell 3 or 7 should be
stimulated) and 3 B (cell 1 or 2 should be stimulated) in our study. The system
of cells 1, 2, 5, 7 in Fig.~1a in \cite{niessen-00} (if the rest of the cells
are removed) corresponds to the topological configuration in Fig.~3~C (cell 5
should be stimulated), 3 D (cell 2 or 7 should be stimulated) and 3 E (cell 1
should be stimulated) in our study. The system of cells 1, 3, 5, 7 in Fig.~1a
in \cite{niessen-00} (if the rest of the cells are removed) corresponds to the
topological configuration in Fig.~3~F (cell 1 should be stimulated) and 3 G
(cell 3, 5 or 7 should be stimulated) and the system of cells in Fig.~9 in
\cite{evans-99} corresponds to the topological configuration in Fig.~3~F (cell
B should be stimulated) and 3 G (cell A, C or D should be stimulated) in our
study. The systems of cells in Fig.~1c and Fig.~2 in \cite{niessen-00} and in
Fig.~11 in \cite{evans-99} correspond to the topological configuration in
Fig.~3~I and 3~J in the present paper. Finally, we believe that it is possible
to extract the configuration shown in Fig.~3~H from a variety of cellular
tissues.

\section{Results}

We first consider the case of two cells. The first cell is stimulated from 0~s
to $t_2=14$~s with $IP_3^{(max)}=4.2~\rm \mu M$ applied at $t_1=4.2$~s (see
Fig.~4). Some $\rm IP_{3}$ from this impulse dissociates as $J_{deg}$ (eq.
(\ref{eq3})). The remainder of $\rm IP_{3}$ diffuses into the second cell
through gap junctions. The concentration of $\rm IP_{3}$, which is received as
the result of all these processes in the first cell, is shown on Fig.~5
($IP_{3(1)}$ from $0~\rm s$ to $14~\rm s$). The concentration of $\rm IP_{3}$
which diffused into the second cell as the result of diffusion through gap
junctions from the first cell is shown in Fig.~5 ($IP_{3(2)}$ from $0~\rm s$ to
$14~\rm s$). The change of $\rm IP_{3}$ in both cells occurs in this period by
large advances, rather than smoothly. This is connected with the fact that the
gap junctional conductance between cells is changed according to $\gamma_{ij}$
(eq. (\ref{eq4})), rather than remaining constant all the time. Thus, as gap
junctions are completely closed at a high concentration of cytosolic $\rm
Ca^{2+}$ in any  two cells, the diffusion of $\rm IP_{3}$ through them
temporarily ceases.

Cytosolic $\rm Ca^{2+}$ oscillations in the model cell  occur at concentrations
of $\rm IP_{3}$ about $1~\rm \mu M$. This concentration of $\rm IP_{3}$  is the
required level in the second cell that occurs at a time later than in the first
cell. Therefore, the cytosolic $\rm Ca^{2+}$  oscillations in the second cell
arise after a certain time delay (see Fig.~6). We find in our model an inverse
dependence of this time delay on the amplitude of the impulse of $\rm IP_{3}$
which is applied to the first cell (see Fig.~6).

For more than two connected cells, our model shows that the propagation of
calcium waves depends on the spatial configuration of cells; this may be
interpreted as the first indication of heterogeneity in intercellular $\rm
Ca^{2+}$ wave propagation. There are three possible configurations of
interactions between three cells (see Fig.~2). If the location of the two
neighboring cells is symmetric with respect to the first cell (Fig.~2~B,~C),
than one can see similar oscillations in  both the second and third cells,
following an initial delay of about $2~\rm s$ from the onset of oscillations in
the first cell. In the case of a linear chain (Fig.~2~A), there is a difference
in the calcium oscillations in the second and third cells. There is not only a
delay time between the onset of calcium oscillations in the first and second
cells, but also between the second and third cells (both delay times are about
$6~\rm s$). The value of this delay time slightly differs from that found in
\cite{sanderson-90}, \cite{sneyd-94}. This  is due to the different choice of
parameter values in the two studies.

We also briefly describe the results for the case of four connected cells. The
results for all configurations of four cells are summarized in Tables 2 -- 4.
We note that we study only in-plane (two-dimensional) cell configurations. It
is possible to distinguish five different configurations of four cells on a
plane. However, for the same topological configuration of cells, the $\rm
Ca^{2+}$ signaling greatly depends on which cell in the structure is
stimulated. Taking into account this fact, we identify ten cell configurations.
As in the case of three cells, the dynamics of $\rm Ca^{2+}$ in cells
symmetrically positioned with respect to the stimulated cell is identical; for
example, see configurations "A", "B", "C", "E", "F", "G" and "H" in Fig.~3.
This holds when all the cells in these configurations are indistinguishable.
Also, this assumption allows us to study the dependence of the intercellular
$\rm Ca^{2+}$ signaling only on the topology of the cell to cell connections in
a given cell configuration.

We now discuss the key features of signaling in four cells configurations. For
example, in the case of configuration "H" (Fig.~3~H) the first cell has two
connections with adjacent cells (there are four connections between cells
altogether) and we find very good calcium wave propagation between cells (see
Fig.~7). Also, the dynamics of $\rm Ca^{2+}$ in cells 2 and 3, which are
positioned symmetrically with respect to the stimulated cell 1, is the same
(see Tables~2 and 4, Fig.~7). The delay time between the onset of $\rm Ca^{2+}$
oscillations in the stimulated cell and the beginning of oscillations in all
other cells in the configuration depends on the ordering number of neighbors of
a given cell. In the case of a direct connection with the first simulated cell
(first order neighbors) the delay time varies from $2~\rm s$ to $4~\rm s$ (see
Table 3). For example, in configuration "I" cell 2 is the first order neighbor
of cell 1. For a second order neighbor, there is one additional cell between
the stimulated cell and a given cell. For example, in configuration "I" cell 3
is the second order neighbor for cell 1. For  second order neighbors, the time
delay for onset of $\rm Ca^{2+}$ oscillation varies from $4~\rm s$ to $11~\rm
s$; this is larger then in the case of the first order neighbors (see Table 3).
In addition, for cells next to the stimulated cell there is a difference in
response depending on the presence or absence of connections with other second
order neighbors (compare cell 3 and cell 4 in configurations "C" and "G",
Tables 2 -- 4).  This suggests a heterogeneity in the $\rm Ca^{2+}$ response in
these cell networks.  Of course, larger cell networks would have to be studied
to strengthen this conclusion. For in-plane configurations of four cells there
is only one neighbor of third order which is present in configuration "I". For
the given parameters there are no oscillations in cell 4 of configuration "I",
as indicated in Tables 2 -- 4 and Fig.~8.

Thus, each cell configuration can be characterized by a definite set
of neighbors of each order; moreover, the propagation of
intercellular $\rm Ca^{2+}$ waves for each configuration depends on
this set. According to the set of neighbors, all cell configurations
can be divided into four classes:
\begin{itemize}
\item one neighbor of first order, one neighbor of second order, one neighbor of third order (configuration "I");
\item one neighbor of first order, two neighbors of second order (configurations "C", "G");
\item two neighbors of first order, one neighbor of second order (configurations "A", "D", "H", "J");
\item three neighbors of first order (configurations "B", "E", "F").
\end{itemize}

A fast decline of intercellular $\rm Ca^{2+}$ waves is characteristic of the
first two classes, whereas a steady expansion of intercellular $\rm Ca^{2+}$
waves is characteristic of the last two classes (see Tables 2 -- 4). For
example, in the case of configuration "C" (Fig.~3~C) the first cell has only
one connection with another cell (there are four connections between all cells,
as in "H") and this results in a decreased range of the calcium wave
propagation from the first cell (see Fig.~7). There is an even stronger
decrease in the range of wave propagation in the case of configuration "I",
which is a linear chain of four cells (see Fig.~8).

We now consider some interesting configurations of four cells in
more detail. Configuration "C" (Fig.~3~C) belongs to the second
class of configurations. In this configuration the
 cytosolic $\rm Ca^{2+}$ displays the typical dynamics of this class, i.e.,
prolonged high-frequency $Ca_{cyt}$ oscillations in the first cell and shorter
$Ca_{cyt}$ oscillations with lower frequency in the other cells. Cells 3 and 4
are equally spaced with respect to the first cell in configuration "C";
therefore their cytosolic $\rm Ca^{2+}$ dynamics is identical. At the same time
the total intensity of oscillations decreases with the distance from the first
cell. Configuration "D" (Fig.~3~D) belongs to the third class in our
classification. The topological difference between configurations "D" and "J"
consists in the presence (configuration "D") and absence (configuration "J") of
a connection between cells 2 and 3. This difference is manifested in $\rm
Ca^{2+}$ dynamics, primarily through small temporary distinctions (see Tables 3
and 4) and also in a weaker response in cell 2 in configuration "D", as
compared with configuration "J" (see Table 4). This is connected with the fact
that cell 2 in configuration "D" exchanges $\rm IP_{3}$ with cell 3 in addition
to cell 1, whereas there is no such exchange in configuration "J". Note that
although in configuration "D", the disposition of cells 2 and 3 in the absence
of cell 4 is symmetrical relative to cell 1 (Fig.~3~D), the cytosolic $\rm
Ca^{2+}$ dynamics in cells 2 and 3 is different (see Fig.~7). This is the
result of the fact that cell 4 is an additional neighbor of cell 3, and an
additional exchange of $\rm IP_{3}$ between these two cells takes place. That
exchange decreases the duration of oscillations in cell 3 in comparison with
cell 2 in configuration "D" (see Fig.~7).

We find one more interesting regularity in the case of different configurations
for four cells. We calculate the sum of the durations of $\rm Ca^{2+}$
oscillations (last column in Table 2) and the sum of the number of peaks of
$\rm Ca^{2+}$ oscillations (last column in Table 4) in all four cells, for each
configuration.  Then we divide this total number of $\rm Ca^{2+}$ oscillations
by the corresponding total duration of $\rm Ca^{2+}$ oscillations for each
configuration and determine the mean frequency of oscillation for each
configuration (Fig.~9). There is obviously a good correlation of this mean
frequency with the classes of configurations described above. In other words,
we find a correlation of each configuration class with a mean frequency for
this class. This regularity allows us to conclude that the topology of the
graph of intercellular connections has a more important influence on
intercellular $\rm Ca^{2+}$ signaling than one might think at first glance.

We find that a variable gap junctional conductance can significantly influence
the range of intercellular $\rm Ca^{2+}$ wave propagation. This is most clear
for the chain of four cells (configuration "I", Fig.~3~I). If the gap
junctional conductance is described by $\gamma_{ij}$ (eq. (\ref{eq4})), $\rm
Ca^{2+}$ oscillations take place only in cells 1, 2 and 3, but the $\rm
Ca^{2+}$ wave does not reach cell 4 (Fig.~8). If the gap junctional conductance
is constant and independent of $Ca_{cyt}$ ($\gamma_{ij}=1$), $\rm Ca^{2+}$
oscillations take place in all four cells of configuration "I" (Fig.~10).
Therefore, even a brief closure of the gap junctions between cells has a strong
effect on the propagation of intercellular $\rm Ca^{2+}$ waves.

We also consider  the effect of varying the parameter $k_G$ on the range of
intercellular $\rm Ca^{2+}$ wave propagation. If one increases  $k_G$,  the gap
junctional conductance $\gamma_{ij}$ decreases. For configuration "I" with
$k_G=0.45~\rm \mu M$, such a decrease of $\gamma_{ij}$ results in an abrupt
limitation of intercellular $\rm Ca^{2+}$ wave propagation. In this case, more
intensive $\rm Ca^{2+}$ oscillations with increased frequency and duration in
comparison with $k_G=0.4~\rm  \mu M$ (see Fig.~8) take place in the first cell.
In the second cell the frequency of $\rm Ca^{2+}$ oscillations became
appreciably smaller in comparison with $k_G=0.4~\rm  \mu M$, whereas in the
third and fourth cells there is only a negligible increase in cytosolic $\rm
Ca^{2+}$ above its baseline value. If one decreases $k_G$, the gap junctional
conductance $\gamma_{ij}$ increases. For configuration "I" with $k_G=0.35~\rm
\mu M$ the corresponding increase in $\gamma_{ij}$ results in an increase in
the range of $\rm Ca^{2+}$ waves and higher frequency oscillations similar to
$\gamma_{ij}=1$ at $k_G=0.4~\rm \mu M$ (Fig.~10).

Finally, we find the following results for all the configurations. First
(denoted as A below), with a constant gap junctional conductance
($\gamma_{ij}=1$), the number of $\rm Ca^{2+}$ oscillations and their total
duration
 increase,  in contrast to the
case with a variable $\gamma_{ij}(Ca_{cyt})$.  Second (denoted as B
below),  several things remain the same whether or not the gap
junction conductance varies with concentration. These include the
response of symmetrically positioned cells, the general dependence
of frequency and duration of $\rm Ca^{2+}$ oscillations in a cell on
the number of its direct neighbors, and the relation between the
time delays from the onset of oscillations in the first cell to the
first oscillations in the other cells of the configuration.

The above-mentioned features are the consequence of $\rm IP_{3}$ diffusion
through the gap junctions. At the same time a variable gap junctional
conductance influences only the absolute  (rather than relative) values of $\rm
IP_{3}$. Thus, independently of whether there is a variable or constant gap
junctional conductance, the qualitative dynamics of cytosolic $\rm Ca^{2+}$
during intercellular signaling for different topological configurations of
cellular connections remain the same. This conclusion remains true if a
different model of intracellular $\rm Ca^{2+}$ is used. Thus, the topology of
cellular connections determines the signaling.

The first features (A) described above are the consequence of the different
absolute values of $\rm IP_3$ which are acquired by each cell in configurations
with open and closed gap junctions. In the case of the features (B), it is not
important how much $\rm IP_3$ has arrived in a cell, since these are relative
characteristics, and only the relative amount of $\rm IP_3$ which moves between
neighboring cells is important. For example, it is important whether the
neighboring cells obtained equal amounts of $\rm IP_3$ or not;  if not, the
difference in the amounts is important. Thus features (A) are absolute
characteristics of the oscillations, and are determined by the behavior of the
gap junctions, whereas features (B) are relative characteristics of the values
in cells (of a given configuration) with respect to each other and are
determined by the topological structure.

Next, we plot in Fig.~11, the same diagram as in Fig.~9, but for the case of
$\gamma_{ij}=1$. We find that we can distinguish only the first (''I'') and the
second (''C'', ''G'') configuration classes, but we cannot distinguish between
the third and fourth classes in our configuration classification. To conclude,
a variable gap junctional conductance has a modulating influence on
intercellular $\rm Ca^{2+}$ signaling by reinforcing the role of the cellular
connection topology, but leaves the qualitative aspects unchanged.

We have also considered $\rm Ca^{2+}$ diffusion through the gap junctions
between cells, in addition to $\rm IP_3$ diffusion between cells. We thus added
an exchange term to eq. (\ref{eq1}) in the form
$k_{gap}\gamma_{ij}(Ca_{cyt(j)}-Ca_{cyt(i)})$, similar to eq. (\ref{eq5}). The
results we obtain allow us to conclude that there are two separate mechanisms
involved in intercellular signaling. First, $\rm IP_3$ diffusion provides the
mechanism for calcium wave propagation between cells, as was proposed in
\cite{sneyd-94}. Second, the diffusion of $\rm Ca^{2+}$, is necessary (in
addition to $\rm IP_3$ diffusion), to synchronize the oscillations among the
neighboring cells. Similar conclusions were drawn by Tsaneva-Atanasova et al.
\cite{tsaneva-05}, which are in complete agreement with this work.

\section{Conclusion}

Intercellular $\rm Ca^{2+}$ signaling is a complex and nonlinear process; its
mechanism depends on the kind of tissue and the method used to stimulate the
cells. In this work we study only plane tissues, cells of which are connected
to each other by gap junctions, such as epithelium and blood vessels. We
suggest that the diffusion of $\rm IP_{3}$ through gap junctions is the main
mechanism for intercellular $\rm Ca^{2+}$ wave propagation in such tissues.
That is, we consider intercellular topological structures which are common to
epithelium and endothelial tissues, but we do not model the details of the
intracellular signaling specific to these cell types.

The intercellular $\rm Ca^{2+}$ wave propagation in airway epithelial cell
cultures arises as the result of mechanical stimulation of a single cell in the
tissue \cite{sanderson-90}. The front of this wave is heterogeneous in
different directions; thus one can observe many kinds of ''branches'', ''paws''
and so on, as well as an oblongness of the front on a whole in some directions
\cite{sanderson-90}. The reasons for such a heterogeneous wave propagation
might include both
 the heterogeneity of the tissue structure (for example, the
different number of gap junctions in the boundaries between adjacent
cells, the different number of neighbors for each cell, the
different shape of cells and so on), and the internal heterogeneity
of the cellular structure (I.V.Dokukina, M.E.Gracheva, E.A.Grachev -
unpublished results). In the current study we show that different
numbers of neighbors results in a heterogeneity of intercellular
$\rm Ca^{2+}$ wave propagation, while varying the gap junctional
conductance amplifies this influence.

We believe that the topology of intercellular connections in tissue
has an essential influence on $\rm Ca^{2+}$ signaling. We
demonstrate this influence for the case of a small (three of four)
number of cells, using the minimal model \cite{goldbeter-90} for the
description of intracellular processes. In addition, we take into
account the dependence of the gap junctional conductance on the
cellular concentration of cytosolic $\rm Ca^{2+}$. Even in the
simplest case of three cells we find differences in $\rm Ca^{2+}$
signaling between the linear chain of three cells (Fig.~2~A) and the
closed chain in which each cell is connected to its adjacent cells
(Fig.~2~C). We find a stronger influence of the topological
structure of intercellular connections on $\rm Ca^{2+}$ signaling
for different configurations of four cells. We classify all possible
plane configurations of four cells in four different categories,
depending on the number of neighbors of a definite order. We find
that the type of intercellular $\rm Ca^{2+}$ signaling in each
configuration depends on the class of this configuration.
Furthermore we find that by knowing the mean frequency of $\rm
Ca^{2+}$ oscillations in any configuration, it is possible to refer
this configuration to one of the four categories.

Our results suggest that during any investigations of intercellular
$\rm Ca^{2+}$ signaling in a particular tissue, it is necessary to
pay attention to the possible influence of the topology of cellular
connections. This is most important for small pieces of tissue,
because in this case the influence of topology is most evident. Note
that we consider only two-dimensional configurations of cells in
this paper. However, it would be interesting to consider the role of
topological structure in three-dimensional tissues, which is a
subject of our future research. Also, Tsaneva-Atanasova et al.
\cite{tsaneva-05} have shown that an analysis of the
point-oscillator model is insufficient to determine which kinds of
oscillations will appear in clusters of pancreatic acinar cells with
realistic geometries. They also find that geometrical factors play a
crucial role in both intra- and intercellular $\rm Ca^{2+}$
signaling. Therefore, it is important to carry out a study of
spatially distributed models of the most interesting topological
configurations of cellular connections.

While modeling the intercellular calcium signaling it is necessary
to recognize that the transition from one cell to tissue is not a
simple quantitative conversion. It is not a transition from one
cell to a chain of cells or some grid of cells; rather, it is a
more complex transition defined by the topology of tissue. We have
shown that the dynamics of tissue in small areas  (and also in
larger regions -- I.V.~Dokukina, A.A.~Tsukanov, M.E.~Gracheva,
E.A.~Grachev, unpublished results) is defined not by a chain or
grid of cells, but by the topology of the connection graph of
cells with each other. It is possible to judge the tissue topology
from the dynamics of signaling; in turn, the tissue topology can
change the intercellular signaling dynamics, enforcing its own
rhythm. Thus, the intercellular dynamics of calcium and the
corresponding topology of cell connections form an unbroken,
complete system.

\section{Acknowledgments}  This research was supported in part by
grants from the National Science Foundation (DMR0302598), and the
G. Harold and Leila Y. Mathers Charitable Foundation.

\newpage

\textbf{Figure Captions}

Fig. 1. Behavior of function
$\gamma_{ij}=\frac{k_{ij}^n}{Ca_{cytMax(ij)}^n+k_{G}^n}$ for different values
of the power $n$. Here $k_{ij}=1~\rm \mu M$ at $n=5$, $k_{ij}=1.84~\rm \mu M$
at $n=3$ and $k_{ij}=3.94~\rm \mu M$ at $n=2$. We represent $\gamma_{ij}$ in
percentile form, or in other words normalize $\gamma_{ij}$ by
$max(\gamma_{ij})$ to obtain the percentile value of $\gamma_{ij}$  from its
maximum. For other parameters see Table \ref{tabl1article}.

Fig. 2. Different structures of connections between three cells.
The grey color of cell 1 means that this cell is stimulated. The
right side shows graphs that correspond to each configuration.

Fig. 3. Different structures of connections between four cells. The grey color
of cell 1 means that this cell is stimulated. The right side shows graphs that
correspond to each cell configuration.

Fig. 4. Schematic representation of stimulus induced $\rm IP_{3}$ as a function
of time.

Fig. 5. Intercellular calcium signaling between two cells based on diffusion of
$\rm IP_{3}$ through gap junctions. Gap junctional conductance depends on
cytosolic $\rm Ca^{2+}$. For all parameters see Table \ref{tabl1article}.

Fig. 6. The dependence of the time delay between initiation of oscillations in
the first (stimulated) and the second cells on the amplitude of the initial
impulse of $\rm IP_{3}$ in the first cell.

Fig. 7. Intercellular calcium signaling between four cells of configurations
"C", "D" and "H" (Fig. 3 C, D and H) based on diffusion of $\rm IP_{3}$ through
gap junctions. Gap junctional conductance depends on cytosolic $\rm Ca^{2+}$.
For all parameters see Table \ref{tabl1article}.

Fig. 8. Intercellular calcium signaling between four cells of configuration "I"
(Fig.~3 I) based on diffusion of $\rm IP_{3}$ through gap junctions. Gap
junctional conductance depends on cytosolic $\rm Ca^{2+}$. For all parameters
see Table \ref{tabl1article}.

Fig. 9. Axis X: all configurations for four cells are separated into four
different categories, taking into account the number of neighbors (see text for
details). Axis Y: the mean frequency of oscillation for all cells of each
configuration ($\rm mean~frequency =
\frac{total~number}{total~duration}~of~all~oscillations$) with
$\gamma_{ij}(Ca_{cyt})$.

Fig. 10. Intercellular calcium signaling between four cells of configuration
"I" (Fig.~3 I) based on diffusion of $\rm IP_{3}$ through gap junctions. Gap
junctional conductance does not depend on cytosolic $\rm Ca^{2+}$. For all
parameters see Table \ref{tabl1article}.

Fig. 11. Axis X: all configurations for four cells are separated into four
different categories, taking into account the number of neighbors (see text for
details). Axis Y: the mean frequency of oscillation for all cells of each
configuration ($\rm mean~frequency =
\frac{total~number}{total~duration}~of~all~oscillations$) with $\gamma_{ij}=1$.

\clearpage

\begin{table}
\caption{Model parameters} \label{tabl1article}
\begin{tabular}{ll}
\\ \hline
{Designation} &{Parameter values}
\\ \hline {$k_{in}$} & {$1~\rm \mu M\cdot s^{-1}$}
\\
{$k_{out}$} & {$6~\rm s^{-1}$}
\\
{$k_{inIP_{3}}$} & {$1~\rm s^{-1}$}
\\
{$k_{pool}$} & {$1~\rm s^{-1}$}
\\
{$k_{rel}$} & {$500~\rm \mu M\cdot s^{-1}$}
\\
{$k_{serca}$} & {$65~\rm \mu M\cdot s^{-1}$}
\\
{$K_{1}$} & {$2~\rm \mu M$}
\\
{$K_{2}$} & {$0.9~\rm \mu M$}
\\
{$K_{3}$} & {$1~\rm \mu M$}
\\
{$V_{p}$} & {$0.01~\rm s^{-1}$}
\\
{$k_{p}$} & {$1~\rm \mu M$}
\\
{$k_{ij}$} & {$1~\rm \mu M$}
\\
{$k_{G}$} & {$0.4~\rm \mu M$}
\\
{$k_{gap}$} & {$0.005~\rm s^{-1}$}
\\
{$k_{rate}$} & {$1~\rm s^{-1}$}
\\\hline
\end{tabular}
\end{table}

\clearpage

\begin{table}
\caption{Duration of $\rm Ca^{2+}$ oscillations (s)} \label{tabl2article}
\begin{tabular}{llllll}
\\ \hline
{Configuration} & {Cell 1} & {Cell 2} & {Cell 3} & {Cell 4} &
{Total}
\\ \hline {$A$} & {$19$} & {$14.5$} & {$14.5$} & {$15.8$} & {$63.8$}
\\
{$B$} & {$18.7$} & {$14.6$} & {$14.6$} & {$14.6$} & {$62.5$}
\\
{$C$} & {$19.7$} & {$15$} & {$10.3$} & {$10.3$} & {$55.3$}
\\
{$D$} & {$17.6$} & {$16.7$} & {$14.6$} & {$13.6$} & {$62.5$}
\\
{$E$} & {$18.7$} & {$14.6$} & {$14.6$} & {$14.6$} & {$62.5$}
\\
{$F$} & {$18.7$} & {$14.6$} & {$14.6$} & {$14.6$} & {$62.5$}
\\
{$G$} & {$19.7$} & {$15$} & {$10.3$} & {$10.3$} & {$55.3$}
\\
{$H$} & {$19$} & {$14.5$} & {$14.5$} & {$15.8$} & {$63.8$}
\\
{$I$} & {$21.4$} & {$15.4$} & {$7.5$} & {--} & {$44.3$}
\\
{$J$} & {$16.9$} & {$18$} & {$16.4$} & {$11.4$} & {$62.7$}
\\ \hline
\end{tabular}
\end{table}

\clearpage

\begin{table}
\caption{Time delay from the beginning of $\rm Ca^{2+}$ oscillations in the
first cell (s)} \label{tabl3article}
\begin{tabular}{llll}
\\ \hline
{Configuration} & {Cell 2} & {Cell 3} & {Cell 4}
\\ \hline {$A$} & {$2.5$} & {$2.5$} & {$4$}
\\
{$B$} & {$2$} & {$2$} & {$2$}
\\
{$C$} & {$4$} & {$8.5$} & {$8.5$}
\\
{$D$} & {$2$} & {$2.5$} & {$5.5$}
\\
{$E$} & {$2$} & {$2$} & {$2$}
\\
{$F$} & {$2$} & {$2$} & {$2$}
\\
{$G$} & {$4$} & {$8.5$} & {$8.5$}
\\
{$H$} & {$2.5$} & {$2.5$} & {$4$}
\\
{$I$} & {$3$} & {$11$} & {--}
\\
{$J$} & {$2$} & {$3$} & {$7.5$}
\\ \hline
\end{tabular}
\end{table}

\clearpage

\begin{table}
\caption{The number of $\rm Ca^{2+}$ oscillations (items)} \label{tabl4article}
\begin{tabular}{llllll}
\\ \hline
{Configuration} & {Cell 1} & {Cell 2} & {Cell 3} & {Cell 4} &
{Total}
\\ \hline {$A$} & {$9$} & {$6$} & {$6$} & {$6$} & {$27$}
\\
{$B$} & {$8$} & {$6$} & {$6$} & {$6$} & {$26$}
\\
{$C$} & {$13$} & {$6$} & {$4$} & {$4$} & {$27$}
\\
{$D$} & {$9$} & {$7$} & {$6$} & {$5$} & {$27$}
\\
{$E$} & {$8$} & {$6$} & {$6$} & {$6$} & {$26$}
\\
{$F$} & {$8$} & {$6$} & {$6$} & {$6$} & {$26$}
\\
{$G$} & {$13$} & {$6$} & {$4$} & {$4$} & {$27$}
\\
{$H$} & {$9$} & {$6$} & {$6$} & {$6$} & {$27$}
\\
{$I$} & {$15$} & {$7$} & {$3$} & {--} & {$25$}
\\
{$J$} & {$9$} & {$8$} & {$6$} & {$4$} & {$27$}
\\ \hline
\end{tabular}
\end{table}

\clearpage

\begin{figure}
    \centering
    \includegraphics[width=1\textwidth,angle=0]{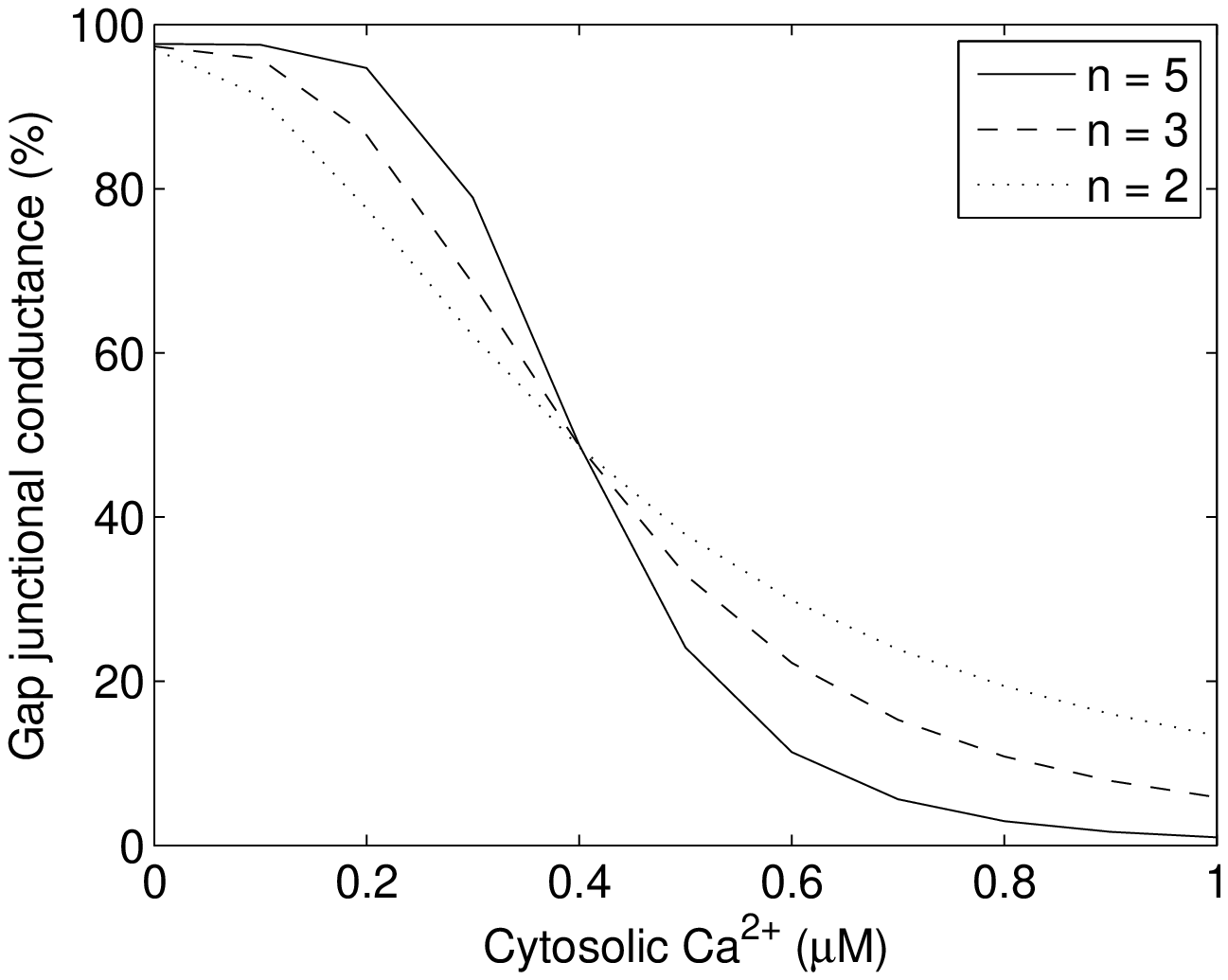}
    \caption{}
    \label{Fig1}
\end{figure}

\clearpage

\begin{figure}
    \centering
    \includegraphics[width=1\textwidth,angle=0]{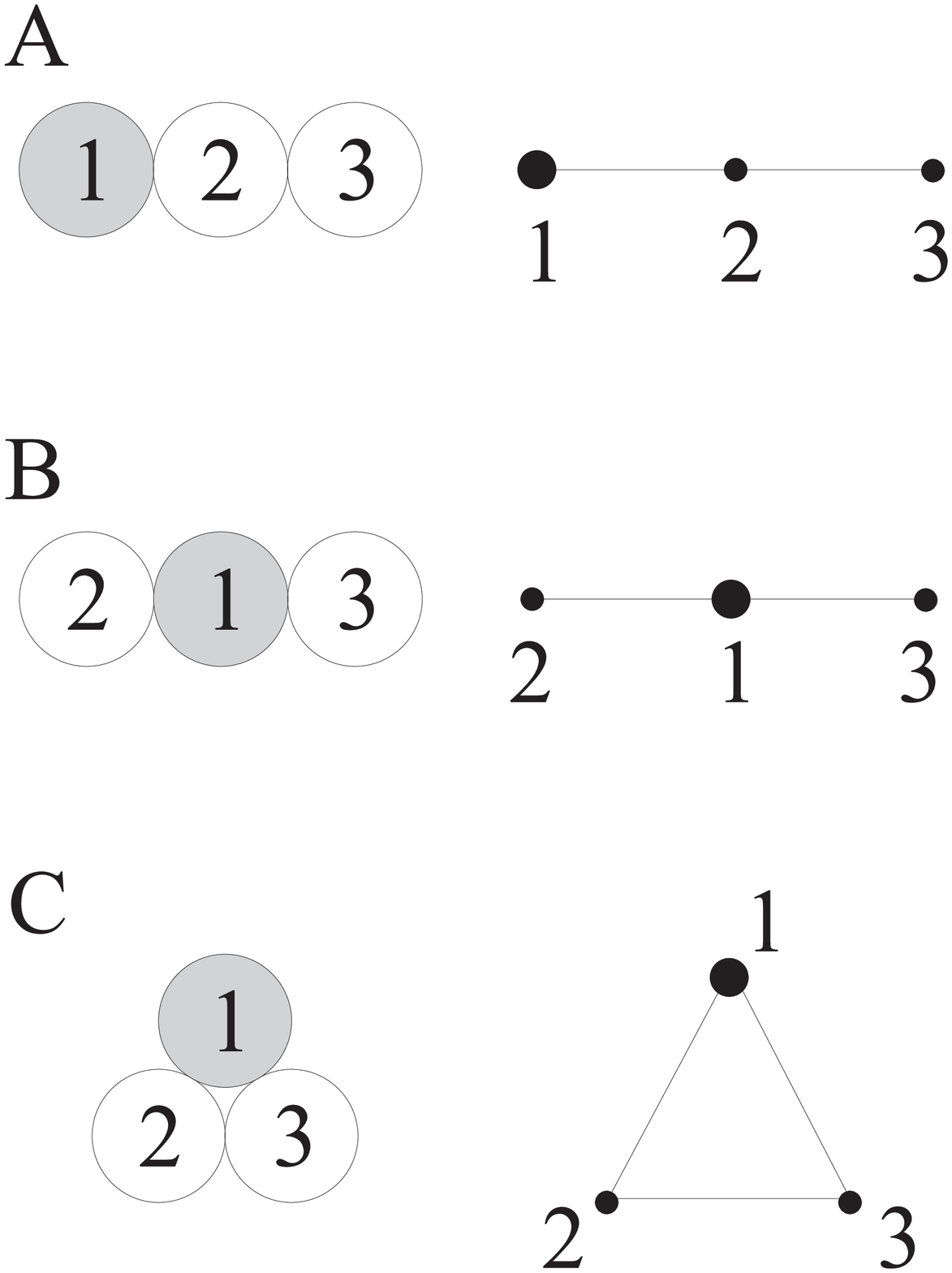}
    \caption{}
    \label{Fig2}
\end{figure}

\clearpage

\begin{figure}
    \centering
    \includegraphics[width=1\textwidth,angle=0]{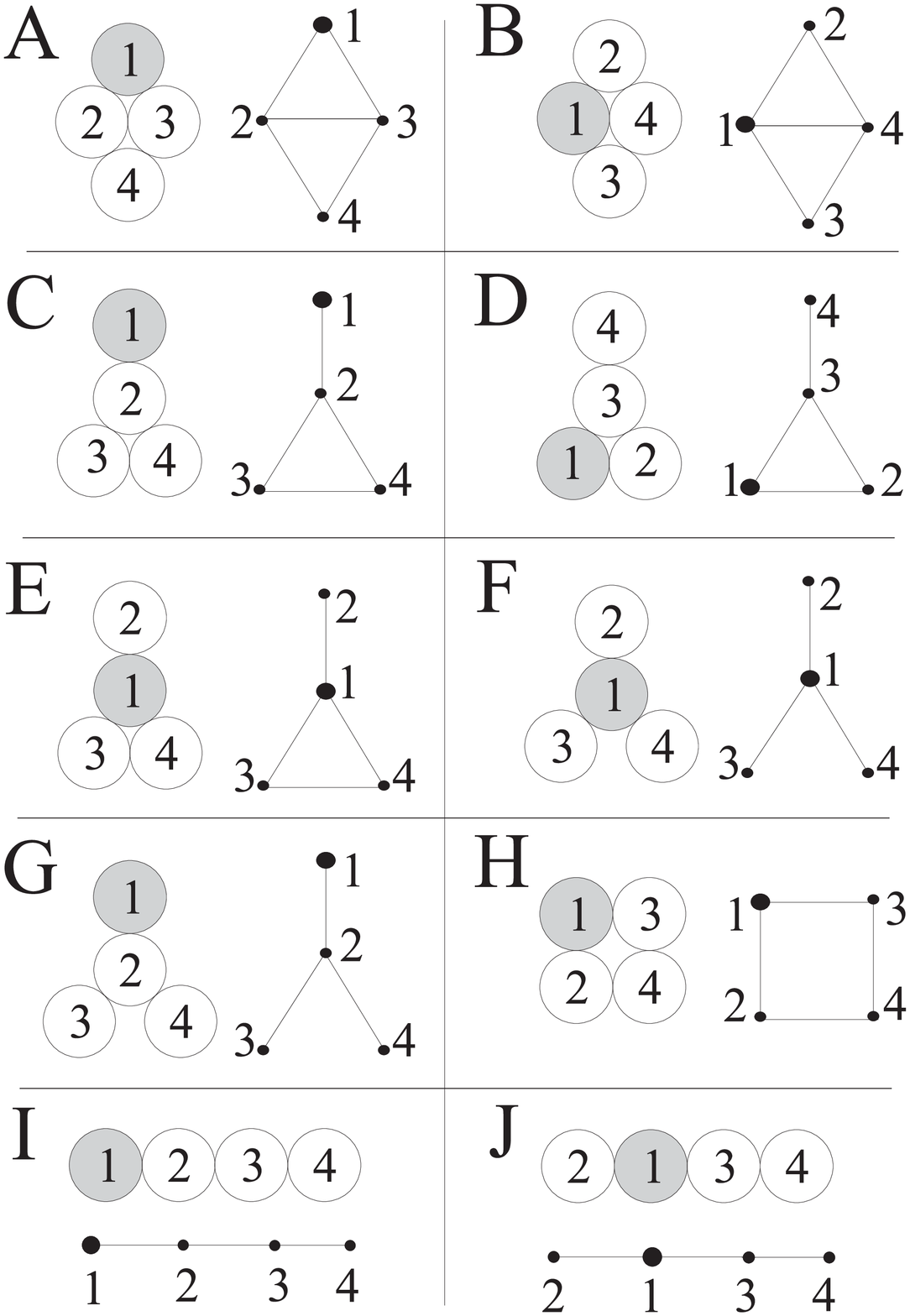}
    \caption{}
    \label{Fig3}
\end{figure}

\clearpage

\begin{figure}
    \centering
    \includegraphics[width=1\textwidth,angle=0]{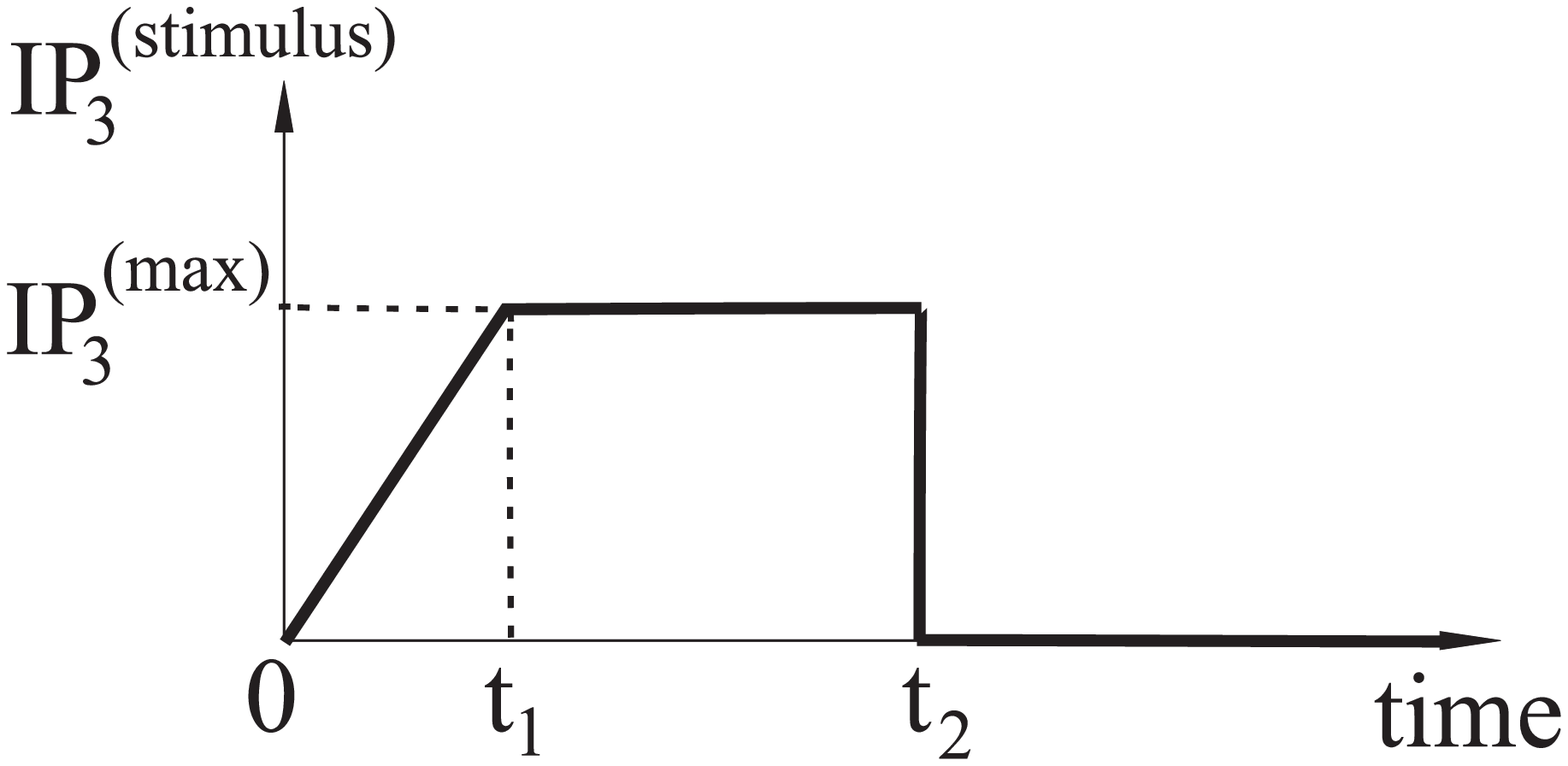}
    \caption{}
    \label{Fig4}
\end{figure}

\clearpage

\begin{figure}
    \centering
    \includegraphics[width=1\textwidth,angle=0]{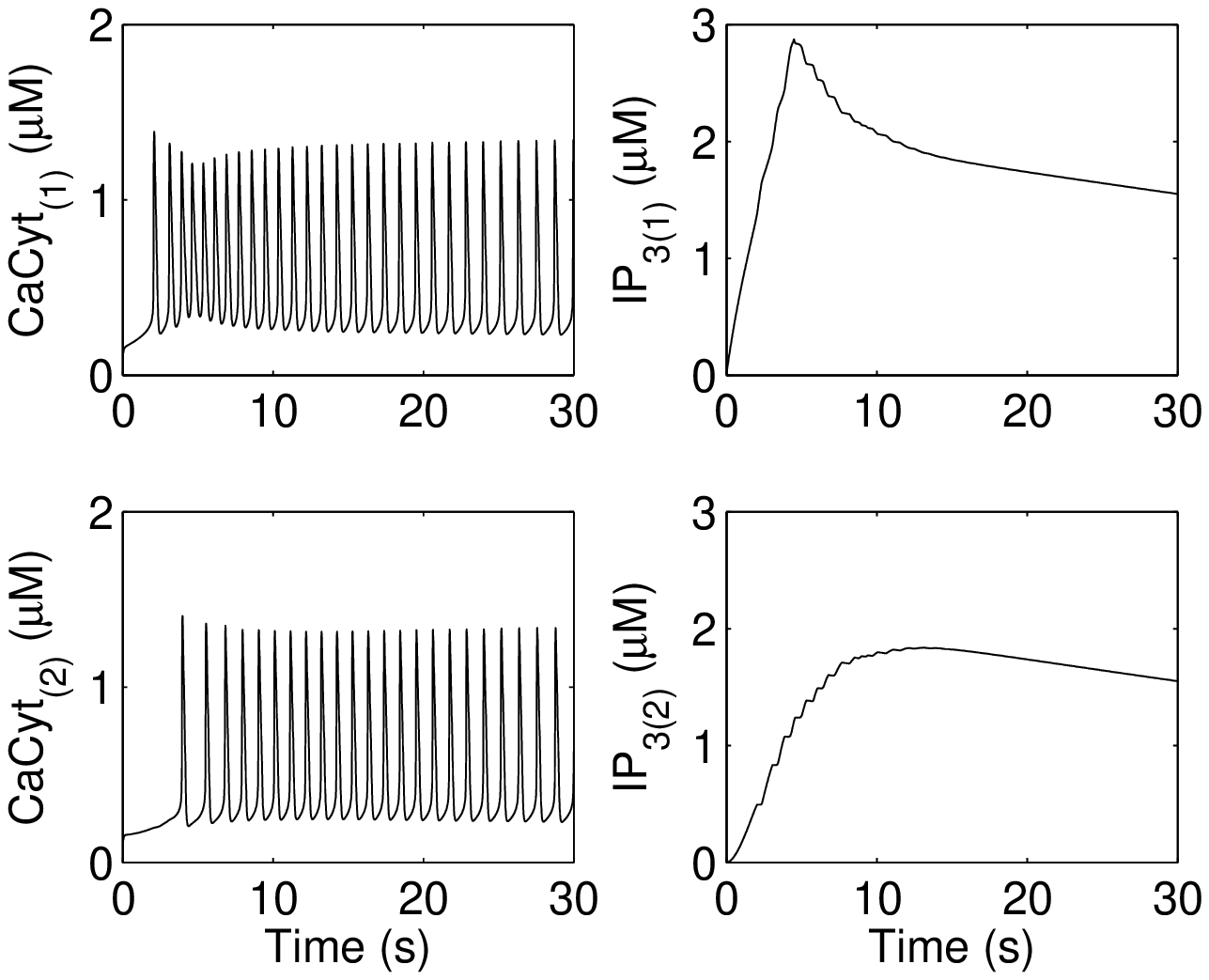}
    \caption{}
    \label{Fig5}
\end{figure}

\clearpage

\begin{figure}
    \centering
    \includegraphics[width=1\textwidth,angle=0]{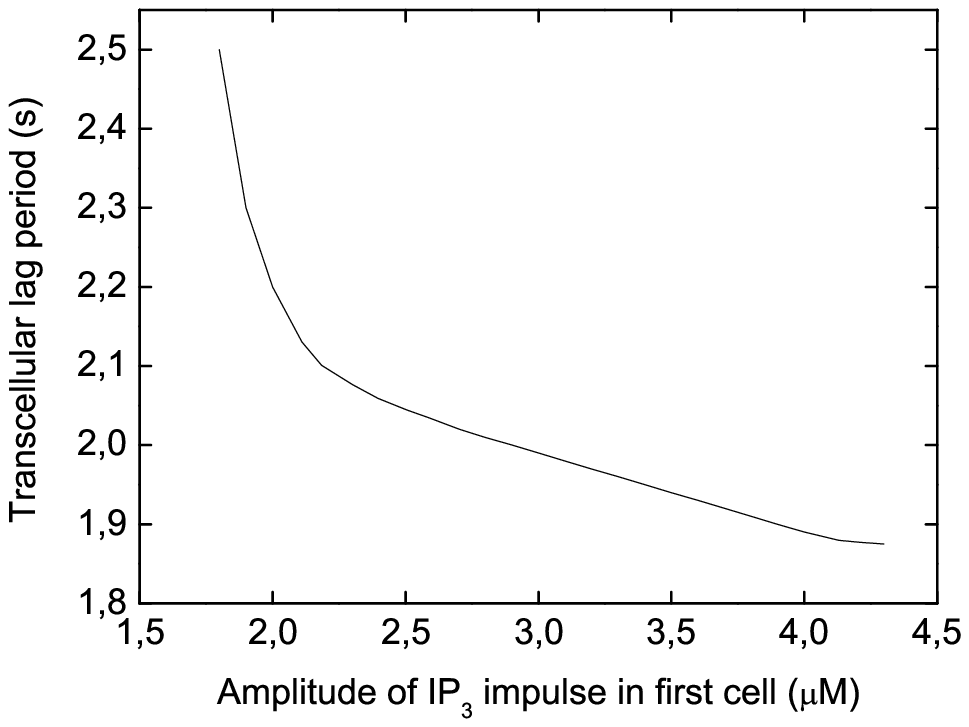}
    \caption{}
    \label{Fig6}
\end{figure}

\clearpage

\begin{figure}
    \centering
    \includegraphics[width=1\textwidth,angle=0]{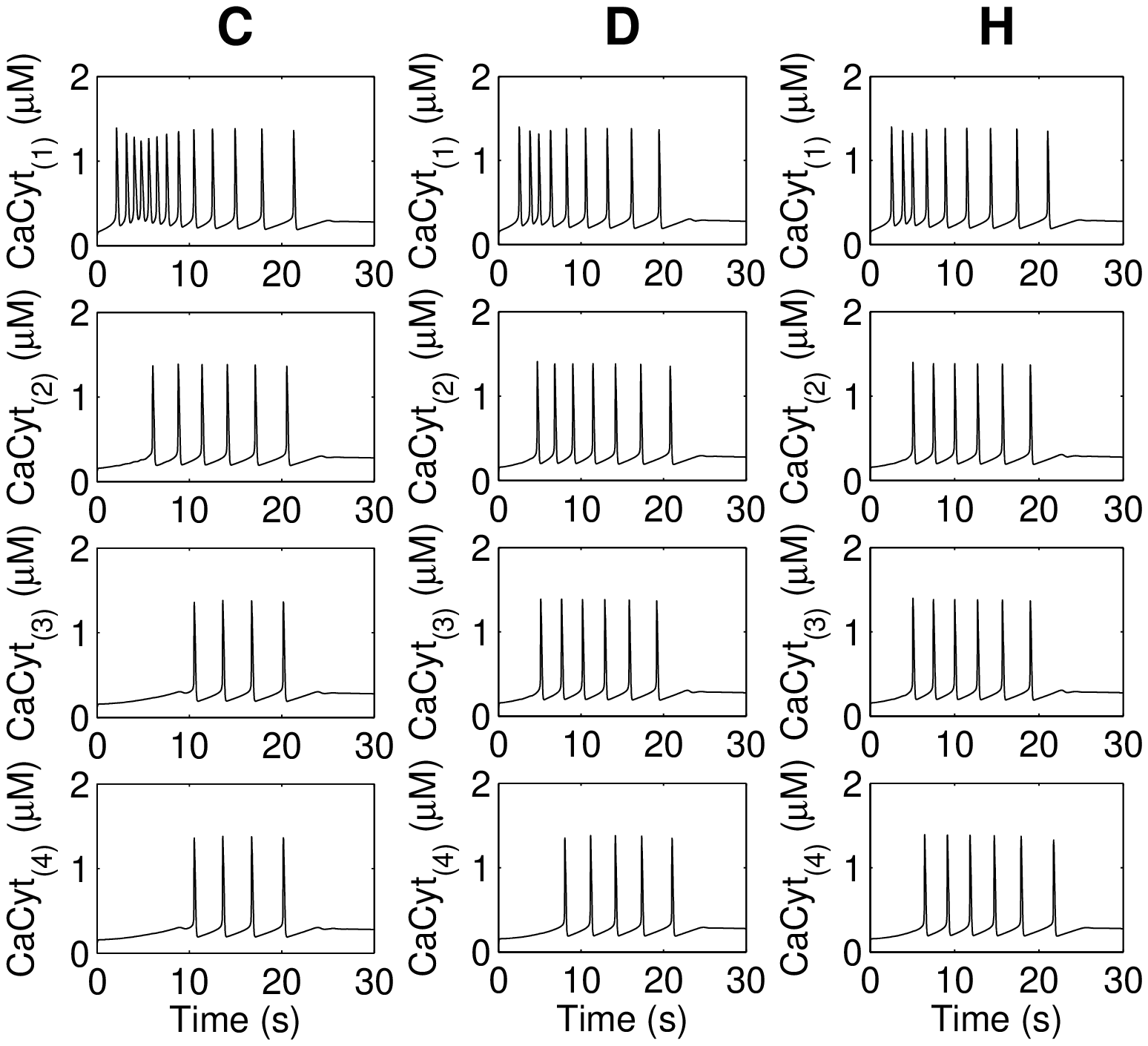}
    \caption{}
    \label{Fig7}
\end{figure}

\clearpage

\begin{figure}
    \centering
    \includegraphics[width=1\textwidth,angle=0]{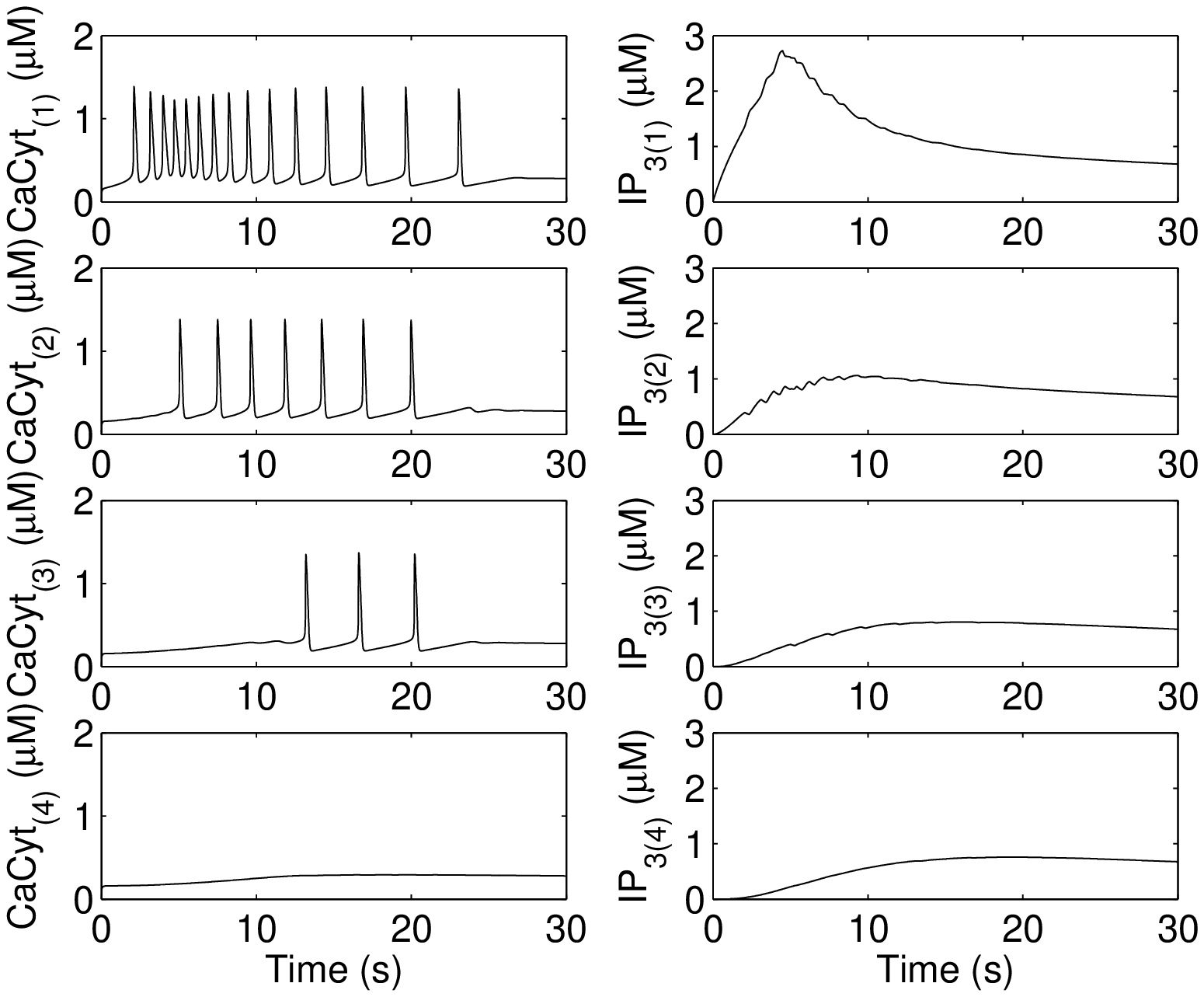}
    \caption{}
    \label{Fig8}
\end{figure}

\clearpage

\begin{figure}
    \centering
    \includegraphics[width=1\textwidth,angle=0]{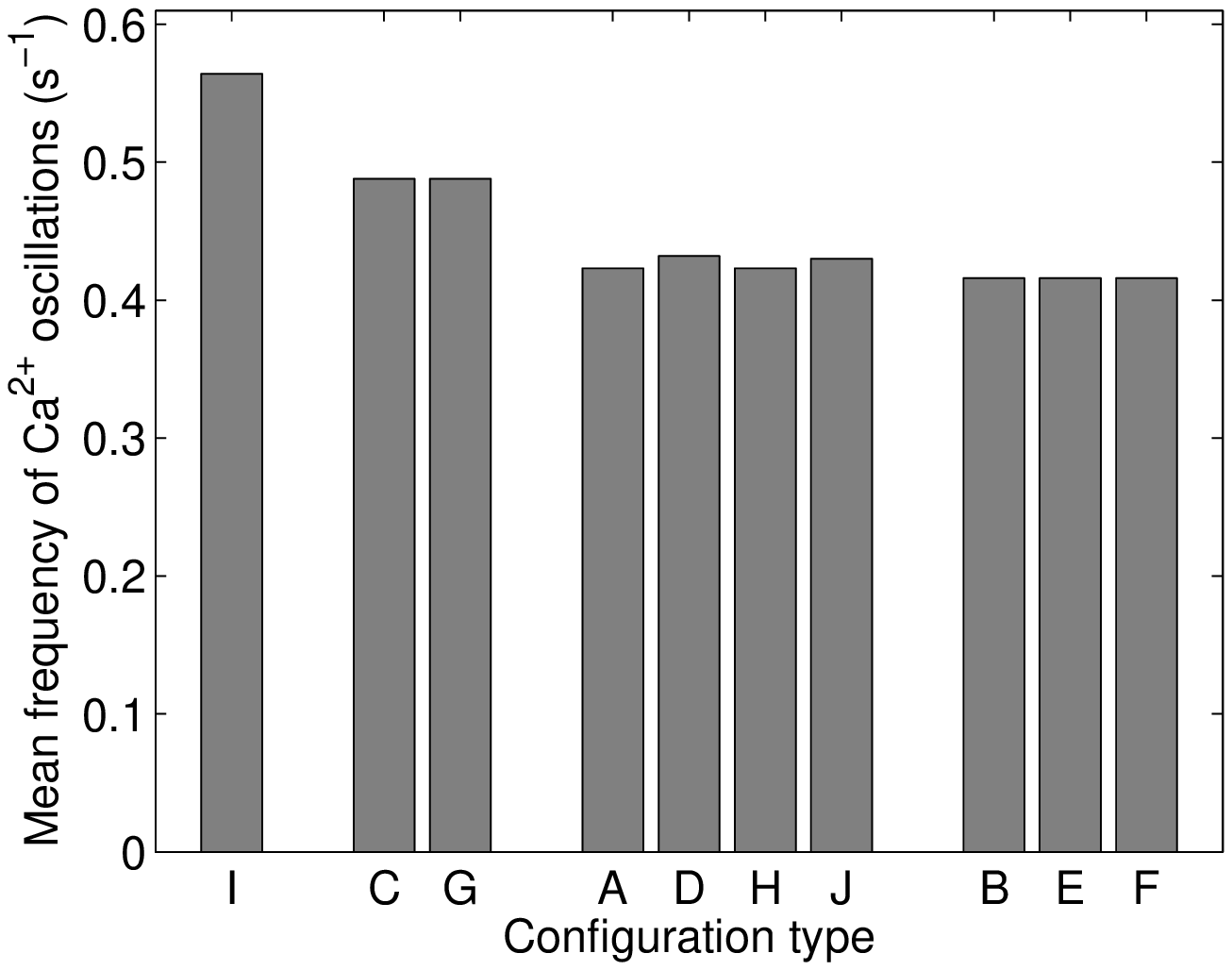}
    \caption{}
    \label{Fig9}
\end{figure}

\clearpage

\begin{figure}
    \centering
    \includegraphics[width=1\textwidth,angle=0]{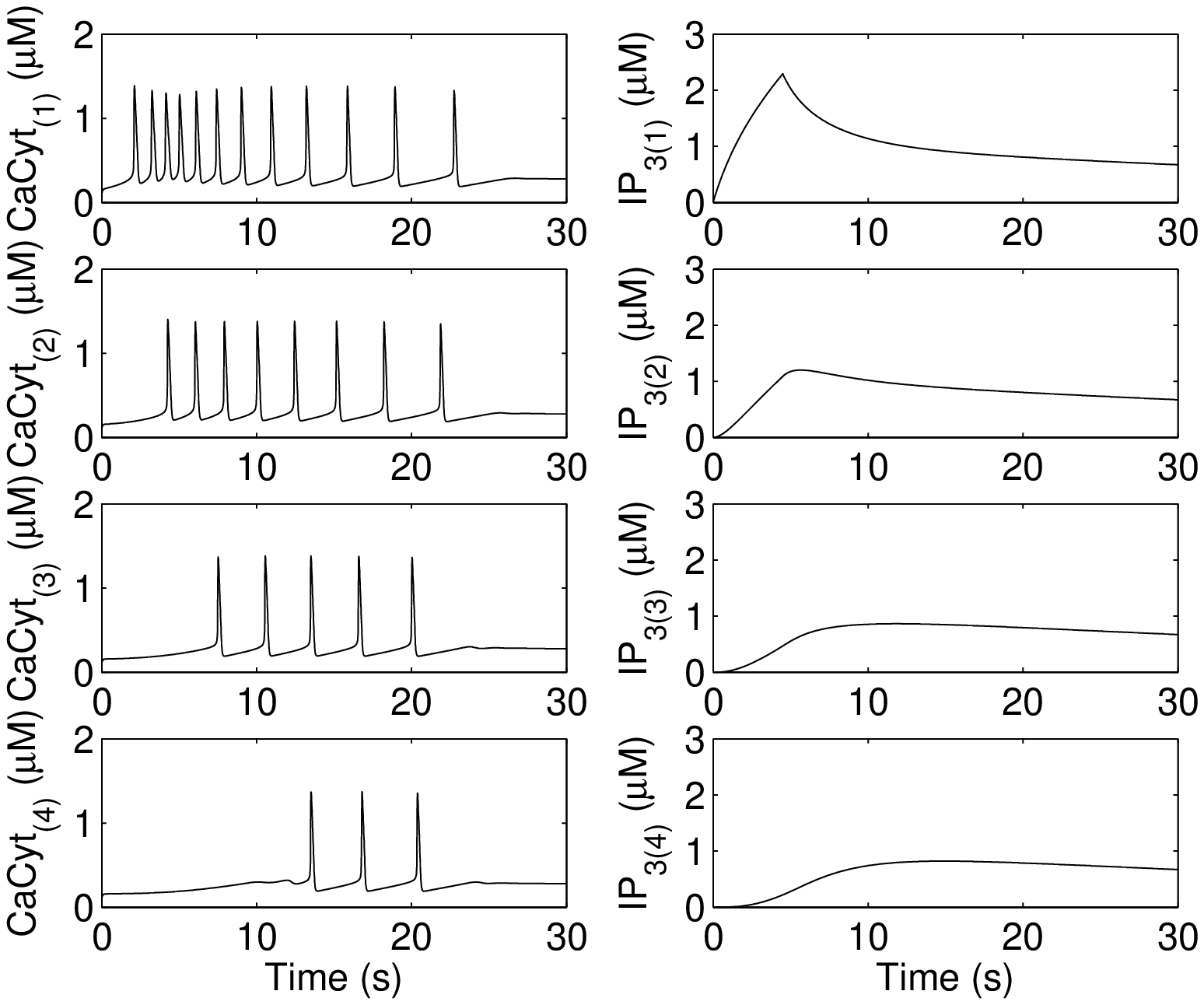}
    \caption{}
    \label{Fig10}
\end{figure}

\clearpage

\begin{figure}
    \centering
    \includegraphics[width=1\textwidth,angle=0]{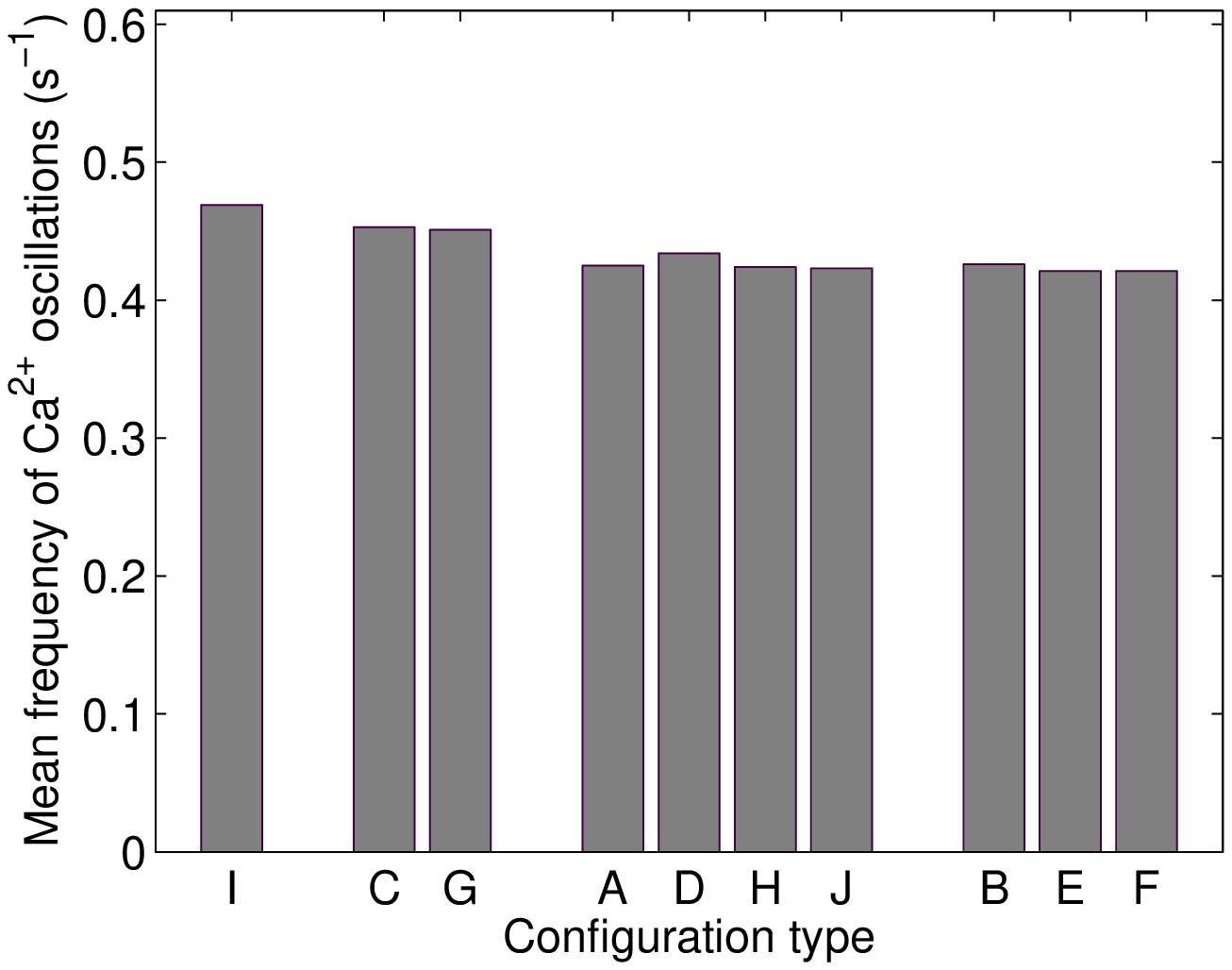}
    \caption{}
    \label{Fig11}
\end{figure}

\end{document}